# A Comprehensive Review of Adversarial Attacks on Machine Learning

Syed Ahmed

syed_ahmed12@infosys.com

Prakhar Mishra

prakhar.mishra02@infosys.com

Bharathi Vokkaliga Ganesh

bharathi_ganesh01@infosys.com

Ravi Anand

ravi.anand06@infosys.com

Sathyanarayana Sampath Kumar

sathyanarayana.k@infosys.com

Bhanuteja Akurathi

bhanuteja.akurathi@infosys.com

Responsible AI Office,

Infosys Limited, Bangalore, India.

## Abstract

This research provides a comprehensive overview of adversarial attacks on AI and ML models, exploring various attack types, techniques, and their potential harms. We also delve into the business implications, mitigation strategies, and future research directions. To gain practical insights, we employ the Adversarial Robustness Toolbox (ART) [1] library to simulate these attacks on real-world use cases, such as self-driving cars. Our goal is to inform practitioners and researchers about the challenges and opportunities in defending AI systems against adversarial threats. By providing a comprehensive comparison of different attack methods, we aim to contribute to the development of more robust and secure AI systems.

## 1. Introduction

*Adversarial attacks* have emerged as a significant threat to the security and reliability of artificial intelligence (AI) and machine learning (ML) systems. These attacks involve crafting malicious inputs that can deceive a model into making incorrect predictions. Research in this area has rapidly expanded in recent years, with a focus on understanding the vulnerabilities of AI models and developing effective defense mechanisms.

*Early research* on adversarial attacks focused on simple methods like adding small perturbations to input data to mislead models. For instance, Szegedy et al. (2014) [2] demonstrated that imperceptible changes to images could cause deep neural networks to misclassify them.

*Subsequent studies* have explored more sophisticated attack techniques, such as gradient-based methods (Goodfellow et al., 2015) [3], optimization-based approaches (Carlini & Wagner, 2017) [4], and transfer-based attacks (Liu et al., 2017) [5]. These attacks have become increasingly effective at evading detection and compromising the accuracy of AI models.

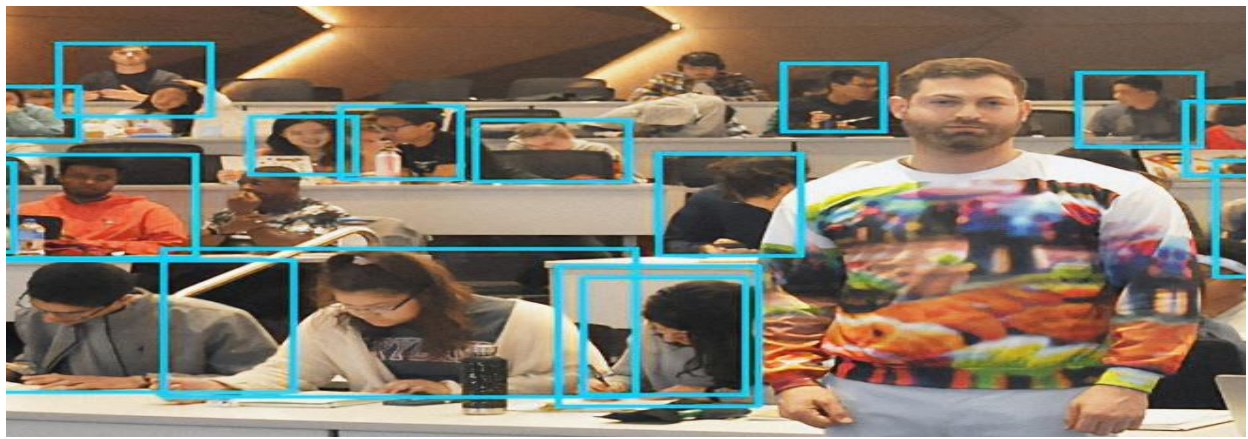

*Figure 1: This visual experiment highlights the vulnerability of AI object detection systems to adversarial manipulation, as evidenced by the ability of a specially designed sweater to evade detection [6].*

*Beyond the technical aspects* of adversarial attacks, researchers have also investigated their potential harms. Adversarial attacks can pose significant risks to security, safety, and privacy. For example, they can be used to compromise autonomous vehicles, manipulate medical diagnosis systems, and steal sensitive information.

### Business Implications of Unmitigated Adversarial Attacks

Adversarial attacks pose significant risks to businesses across various industries. If left unmitigated, these attacks can lead to severe financial losses, reputational damage, and regulatory penalties.

### Financial Losses

- *Direct financial losses***:** Adversarial attacks can result in direct financial losses due to fraudulent transactions, data breaches, and service disruptions. For example, a cyberattack on a financial institution could lead to unauthorized fund transfers and customer data theft.

- *Loss of customer trust***:** Adversarial attacks that compromise customer data or lead to service disruptions can erode customer trust, resulting in decreased revenue and market share.

### Reputational Damage

- *Negative publicity***:** Businesses that are victims of adversarial attacks can suffer significant reputational damage. Negative media coverage and public scrutiny can tarnish a company's brand and lead to customer churn.

- *Regulatory fines***:** In many cases, businesses that fail to adequately protect against adversarial attacks may face regulatory fines and penalties.

### Regulatory Penalties

- *Non-compliance***:** Adversarial attacks can expose businesses to regulatory non-compliance, leading to fines and legal action. For example, companies that fail to comply with data protection regulations like GDPR or CCPA may face hefty penalties.

- *Loss of licenses***:** In some cases, severe breaches of security or privacy can result in the loss of business licenses or certifications.

### Industry-Specific Risks

- *Healthcare*: Adversarial attacks on medical imaging systems could lead to misdiagnosis and incorrect treatment, resulting in patient harm and legal liabilities.

- *Finance*: Financial institutions are particularly vulnerable to adversarial attacks, which can lead to fraud, identity theft, and market manipulation.

- *Autonomous systems*: Adversarial attacks on autonomous vehicles or industrial control systems could have catastrophic consequences, such as accidents or disruptions to critical infrastructure.

### Case Studies

- **Tesla Autopilot Attack***: In 2017, researchers exposed vulnerabilities in Tesla's Autopilot system by placing adversarial stickers on road signs, deceiving the system into misinterpreting traffic signals. These stickers could cause the Autopilot to confuse a stop sign with a yield sign or ignore it altogether, posing significant risks in real-world driving scenarios. This attack underscores the dangers of adversarial machine learning in safety-critical applications and highlights the urgent need for robust defenses against such vulnerabilities [7].

- **The Waymo incident 640** (2023-12-11) involved two collisions between a Waymo vehicle and a towed pickup truck that was improperly angled across a traffic lane. This incident serves as a stark reminder of the potential vulnerabilities of autonomous vehicles, even in seemingly straightforward scenarios**.** While the specific causes of the collisions were related to the vehicle's perception system, the incident underscores the importance of robust safety measures to prevent accidents. If autonomous vehicles are susceptible to errors, as demonstrated by the Waymo case, the introduction of adversarial attacks could amplify these risks significantly**.** Adversarial attacks involve malicious actors deliberately manipulating the vehicle's environment or systems to cause it to behave unexpectedly or dangerously. For instance, an attacker might use camouflage or decoys to confuse the
vehicle's sensors, or they could exploit vulnerabilities in the vehicle's software to override its decision-making processes.

*In response* to the growing threat of adversarial attacks, researchers have developed various defense mechanisms. These include adversarial training (Madry et al., 2019) [8], certified robustness (Wong et al., 2018) [9], and adversarial detection (Ma et al., 2019) [10]. However, the development of new attack techniques continues to outpace defensive measures, making it an ongoing arms race.

*While significant* progress has been made in understanding and mitigating adversarial attacks, several challenges remain. These include the difficulty of detecting and defending against stealthy attacks, the computational cost of robust training, and the generalization of defenses to different models and datasets.

*In conclusion* Adversarial attacks pose a significant threat to the security and reliability of AI and ML systems, potentially compromising their applications in critical domains such as healthcare, finance, and autonomous vehicles. ongoing research aims to comprehensively evaluate the vulnerabilities of widely adopted machine learning models to adversarial attacks. By analyzing factors like attack efficiency, perturbation magnitude, and attack diversity, we seek to identify the most effective attack strategies and inform the development of robust defense mechanisms. Ultimately, this research will contribute to enhancing the security and trustworthiness of AI and ML systems in the face of emerging threats.

In this research, we conducted experiments using two types of machine learning models: one focused on object detection and the other on image classification. Our objective is to apply various adversarial attacks on these models to determine which attacks are more effective and which require greater attention.

After successfully applying these attacks, our goal will be to identify suitable defenses that can enhance the robustness of these models. We will focus on implementing defenses that have a lower computational cost, allowing us to effectively counter these attacks.

Finally, we will present the results and findings from our experiments, highlighting the performance of different attacks and the efficacy of the proposed defenses.

## 2. Experimental Setup

In this research, we combined two experimental setups focusing on different machine learning use cases: object detection and image classification.

For the first use case, we utilized the **DETR-ResNet50 ([facebook/detr-resnet-50 · Hugging Face](#))** model, available on **Hugging Face**. This model was chosen due to its widespread acceptance and demonstrated an AP (average precision) of **42.0** on COCO 2017 validation. We employed a **self-driving car** ([Self-Driving Cars | Kaggle](#)) dataset sourced from **Kaggle**, which contains a total of 22,200 samples. To ensure a focused analysis aligned with our objectives, we randomly selected 100 representative samples for this study. The implementation of adversarial attacks was conducted using the latest version of Python along with the **Adversarial Robustness Toolbox** (ART) (version 1.18.1). Various attacks from the ART [1] library were selected for their relevance to our context. Each attack was initially applied using default parameters, followed by manual adjustments to achieve misclassification with minimal perturbations. The findings and results for each attack, including the selected parameters and their effectiveness, are detailed below.

For the second use case, we employed a **convolutional neural network** (CNN) model trained on a **Bone Break Classification Image ([Bone Break Classification Image Dataset (kaggle.com)](#))** Dataset, which distinguishes between 10 different types of bone fractures. The model achieved an accuracy of approximately 99%. The dataset for this experiment was sourced from **Kaggle**, specifically the Bone Break Classification Image Dataset. Like the first use case, we randomly selected 100 representative samples for focused analysis. The implementation of adversarial attacks followed the same methodology, utilizing Python and the ART [1]toolbox (version 1.18.1). Various attacks were applied with both default and manually adjusted parameters to assess their impact on model performance. The findings and results for this use case, including attack parameters and effectiveness, are also detailed below.

By presenting both experiments together, we aim to provide a comprehensive understanding of adversarial attacks across different machine learning models and use cases.

<div align="center">

**Attacks on Object Detection Use Case**

</div>

1. **FGSM**

   **Definition**: It is an adversarial attack that operates in both targeted and untargeted settings, controlling the $l_1$, $l_2$, or $l_\infty$ norm of the adversarial perturbation. For the targeted case with the $\infty$ norm, the adversarial perturbation is defined as:

   $$\psi(x,y) = -e \cdot \text{sign}(\nabla x L(x,y)) \psi(x,y) = -e \cdot \text{sign}(\nabla x L(x,y))$$

   where $e > 0 e > 0$ is the attack strength and $y y$ is the target class specified by the attacker. The adversarial sample is computed as:

   $$xadv = \text{clip}(x + \psi(x,y), xmin, xmax) xadv = \text{clip}(x + \psi(x,y), xmin, xmax)$$

   In its untargeted form, the FGSM attack is represented as:

   $$\rho(x) = -\psi(x, C(x)) \rho(x) = -\psi(x, C(x))$$

The strength of the FGSM attack depends on the choice of the parameter e*e*. For a comprehensive description of this method and its implementation, refer to the Adversarial Robustness Toolbox v1.0.0 research paper [13].

Having defined the Fast Gradient Sign Method (FGSM) [11]and its operation, we implemented the attack using the **Adversarial Robustness Toolbox (ART)**. The library provides a straightforward interface for executing FGSM, allowing us to leverage its built-in functions for generating adversarial samples efficiently.

**Parameters of the Attack:**

- estimator – A trained classifier.
- norm – The norm of the adversarial perturbation. Possible values: "inf", np.inf, 1 or 2.
- eps – Attack step size (input variation).
- eps_step – Step size of input variation for minimal perturbation computation.
- targeted (bool) – Indicates whether the attack is targeted (True) or untargeted (False)
- num_random_init (int) – Number of random initialisations within the epsilon ball. For random_init=0 starting at the original input.
- batch_size (int) – Size of the batch on which adversarial samples are generated.
- minimal (bool) – Indicates if computing the minimal perturbation (True). If True, also define *eps_step* for the step size and eps for the maximum perturbation.
- summary_writer – Activate summary writer for TensorBoard. Default is *False* and deactivated summary writer. If *True* save runs/CURRENT_DATETIME_HOSTNAME in current directory. If of type *str* save in path. If of type *SummaryWriter* apply provided custom summary writer. Use hierarchical folder structure to compare between runs easily. e.g. pass in 'runs/exp1', 'runs/exp2', etc. for each new experiment to compare across them.

We executed the FGSM attack using the default parameter values provided by the Adversarial Robustness Toolbox (ART). After running the attack with these settings, we observed the initial results. To further refine our approach, we then modified the default parameters, aiming to achieve misclassification with minimal perturbations.

The following table outlines the default and modified parameters used in our ART experiments.

| Attack Parameter | Default Value | Manual Value |
|---|---|---|
| estimator | PyTorch Detector Transformer | PyTorch Detector Transformer |
| norm | Inf | Inf |
| eps | 0.3 | 0.1 |
| epsilon_step | 0.1 | 0.0005 |
| targeted | False | False |
| num_random_int | 0 | 0 |
| Batch_size | 32 | 32 |
| minimal | False | False |
| Summary_writer | False | False |

The effectiveness of these adjustments is demonstrated in the images below, showcasing the results of both the initial and modified attacks.

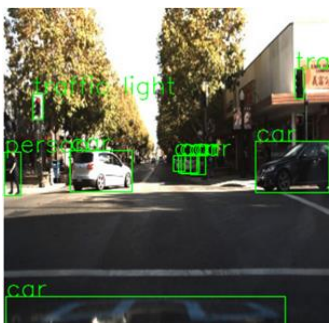

*Original image with prediction.*

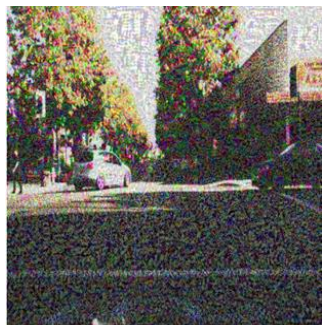

*Adversarial Image generated using FGSM attack by passing default parameters*

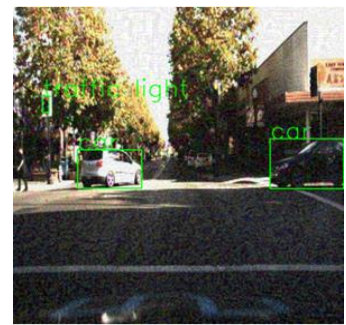

*Adversarial image generated using FGSM attack by passing altered parameters.*

In this example, we demonstrate the generation of adversarial images from an original image. The original image is accurately classified by the model, recognizing objects like cars, traffic lights, and pedestrians. In contrast, both default and manual parameter-based adversarial images successfully mislead the model, resulting in either no object detection or incorrect identifications. Notably, the default parameter-generated image displays more pronounced visual noise, making it visibly distinct from the original.

**Attack Effectiveness:**

The FSGM attack was applied to 100 sample images. The mean perturbation introduced was 0.0916007. Notably, this level of perturbation was sufficient to successfully generate adversarial samples for almost all 100 images, resulting in a 99% attack success rate.

## 2. PGD

Projected Gradient Descent [11] is also an iterative extension of FGSM and very similar to BIM. The main difference with BIM resides in the fact that PGD projects the attack result back on the e-norm ball around the original input at each iteration of the attack.

**Parameters of the Attack:**

- estimator – An trained estimator.
- norm – The norm of the adversarial perturbation supporting "inf", np.inf, 1 or 2.
- eps – Maximum perturbation that the attacker can introduce.
- eps_step – Attack step size (input variation) at each iteration.
- random_eps (bool) – When True, epsilon is drawn randomly from truncated normal distribution. The literature suggests this for FGSM based training to generalize across different epsilons. eps_step is modified to preserve the ratio of eps / eps_step. The effectiveness of this method with PGD is untested (https://arxiv.org/pdf/1611.01236.pdf).
- decay – Decay factor for accumulating the velocity vector when using momentum.

- max_iter (int) – The maximum number of iterations.
- targeted (bool) – Indicates whether the attack is targeted (True) or untargeted (False).
- num_random_init (int) – Number of random initialisations within the epsilon ball. For num_random_init=0 starting at the original input.
- batch_size (int) – Size of the batch on which adversarial samples are generated.
- summary_writer – Activate summary writer for TensorBoard. Default is *False* and deactivated summary writer. If *True* save runs/CURRENT_DATETIME_HOSTNAME in current directory. If of type *str* save in path. If of type *SummaryWriter* apply provided custom summary writer. Use hierarchical folder structure to compare between runs easily. e.g. pass in 'runs/exp1', 'runs/exp2', etc. for each new experiment to compare across them.
- verbose (bool) – Show progress bars.

The following table outlines the default and modified parameters used in our ART experiments.

| Attack Parameter | Default Value | Manual Value |
|---|---|---|
| Estimator | OBJECT_DETECTOR_TYPE | PyTorch Detector Transformer |
| norm | *inf* | *inf* |
| eps | *0.3* | 0.1 |
| eps_step | *0.1* | *0.1* |
| random_eps | *False* | *False* |
| decay | *None* | *None* |
| max_iter | 100 | 100 |
| targeted | *False* | *False* |
| num_random_init | 0 | 0 |
| batch_size | 32 | 32 |
| summary_writer | *False* | *False* |
| verbose | *True* | *True* |

The effectiveness of these adjustments is demonstrated in the images below, showcasing the results of both the initial and modified attacks.

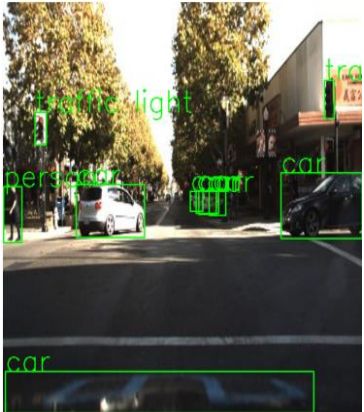

*Original image with prediction.*

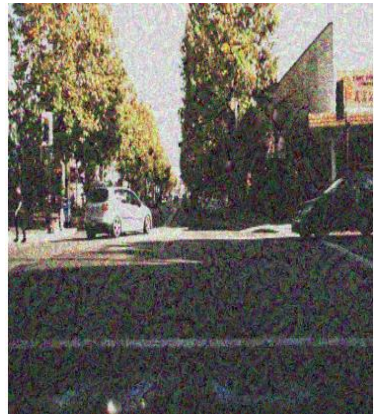

*Adversarial Image generated using PGD attack by passing default parameters.*

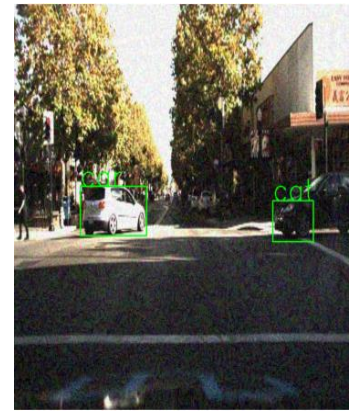

*Adversarial image generated using PGD attack by passing altered parameters.*

In this example, we can clearly observe how we generated an adversarial image from the original. The model effectively classifies objects in the original image, correctly identifying items such as cars, traffic lights, and pedestrians. However, in the adversarial image, the model fails to detect all objects, and those that are detected are misclassified; for instance, a car is incorrectly identified as a cat. Notably, the default parameter-generated image displays more pronounced visual noise, making it visibly distinct from the original.

**Attack Effectiveness:**

The PGD attack was applied to 100 sample images. The mean perturbation introduced was 0.063883. Notably, this level of perturbation was sufficient to successfully generate adversarial samples for all 100 images, resulting in a 100% attack success rate.

3. **Basic Iterative Method**

The Basic Iterative Method (BIM) [11] is an iterative adversarial attack that applies small perturbations to an input in multiple steps, using gradient information to maximize the loss function and effectively misclassify the input.

**Parameters of the Attack:**

- estimator – An trained classifier.
- eps – Maximum perturbation that the attacker can introduce.
- eps_step – Attack step size (input variation) at each iteration.
- max_iter (int) – The maximum number of iterations.
- targeted (bool) – Indicates whether the attack is targeted (True) or untargeted (False).
- batch_size (int) – Size of the batch on which adversarial samples are generated.
- verbose (bool) – Show progress bars.

The following table outlines the default and modified parameters used in our ART experiments.

| Attack Parameter | Default Value | Manual Value |
|---|---|---|
| Estimator | | PyTorch Detector Transformer |
| eps | 0.3 | 0.1 |
| eps_step | 0.1 | 0.05 |
| max_iter | 100 | 100 |
| targeted | False | False |
| batch_size | 32 | 32 |
| verbose | True | True |

The effectiveness of these adjustments is demonstrated in the images below, showcasing the results of both the initial and modified attacks.

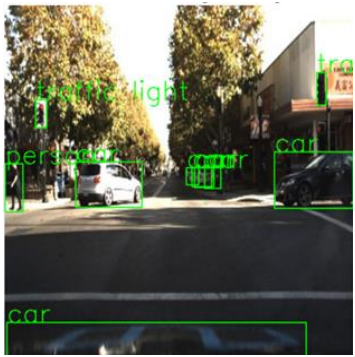

*Original image with prediction.*

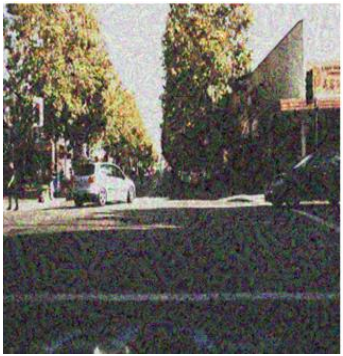

*Adversarial Image generated using BIM attack by passing default parameters.*

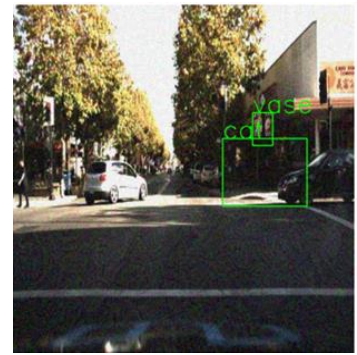

*Adversarial image generated using BIM attack by passing altered parameters*

In this example, we can clearly observe how we generated an adversarial image from the original image. If we compare both adversarial images, then we can say that adversarial image generated by passing altered parameters has less noise in-comparison to adversarial image generated by passing default parameters. The model effectively classifies objects in the original image, correctly identifying items such as cars, traffic lights, and pedestrians. However, in the adversarial image, the model fails to detect all objects, and those that are detected are misclassified; for instance, a car is incorrectly identified as a cat.

**Attack Effectiveness**:

The BIM attack was applied to 100 sample images. The mean perturbation introduced was 0.05783737. Notably, this level of perturbation was sufficient to successfully generate adversarial samples for all 100 images, resulting in a 100% attack success rate.

**Evaluation of Adversarial Attack Efficacy for Object Detection Use case**

| Attack | Mean Perturbation |
|---|---|
| **FGSM** | 0.0916007 |
| **PGD** | 0.063883 |
| **Basic Iterative Method** | 0.05783737 |

To quantify the subtlety of adversarial attacks, we measure mean perturbation, which reflects the average magnitude of pixel modifications needed to induce misclassification.

While the mean perturbation values, representing the average pixel-wise alteration for adversarial attacks, are relatively small for all three methods (BIM, PGD, and FGSM), our experiments reveal a slight advantage for BIM in terms of stealthiness.

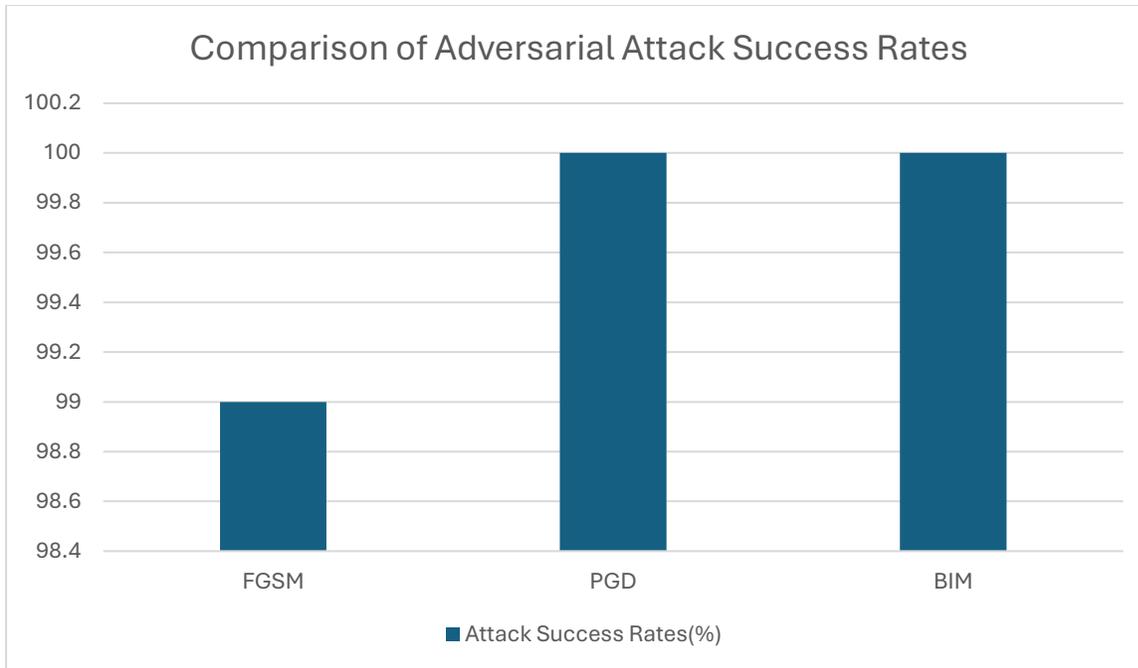

## Conclusion for attacks on Object Detection use case

By employing the ART library, we have shown that object detection models are susceptible to adversarial attacks. Each of the three tested attacks (BIM, PGD, and FGSM) effectively produced adversarial examples that caused the model to misclassify objects. This underscores the importance of developing robust defenses against adversarial attacks, particularly for models deployed in real-world scenarios with potential security implications, such as autonomous driving systems.

To address the vulnerabilities exposed by these attacks, we will discuss potential defense mechanisms in a dedicated section later in this paper.

## Attacks on Bone Break Use Case

### Basic Iterative Method

Definition: The Basic Iterative Method (BIM) [11]is an iterative adversarial attack that applies small perturbations to an input in multiple steps, using gradient information to maximize the loss function and effectively misclassify the input.

Parameters of the Attack:

- estimator – An trained classifier.
- eps – Maximum perturbation that the attacker can introduce.
- eps_step – Attack step size (input variation) at each iteration.
- max_iter (int) – The maximum number of iterations.
- targeted (bool) – Indicates whether the attack is targeted (True) or untargeted (False).
- batch_size (int) – Size of the batch on which adversarial samples are generated.
- verbose (bool) – Show progress bars.

The following table outlines the default and modified parameters used in our ART experiments.

| Attack Parameter | Default Value | Manual Value |
| --- | --- | --- |
| Estimator | *CLASSIFIER_LOSS_GRADIENTS_TYPE* | TensorFlowV2Classifier |
| eps | 0.3 | 1.0 |
| eps_step | 0.1 | 0.005 |
| max_iter | 100 | 10 |
| targeted | False | False |
| batch_size | 32 | 32 |
| verbose | True | True |

The effectiveness of these adjustments is demonstrated in the images below, showcasing the results of both the initial and modified attacks.

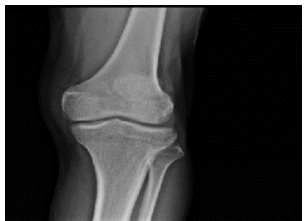

*Original Image*

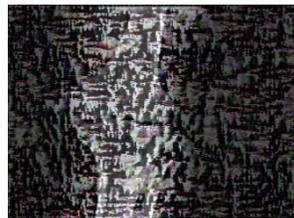

*Adversarial Image generated using BIM attack by passing default parameters.*

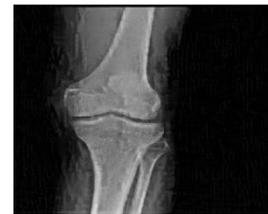

*Adversarial image generated using BIM attack by passing altered parameters.*

Attack Effectiveness:

The BIM attack was applied to 100 sample images. The above adversarial image on default parameters clearly depicts, there is so much noise in-comparison with adversarial image on altered parameters. The mean perturbation introduced was 0.018666986. Notably, this level of perturbation was sufficient to successfully generate adversarial samples for almost all 100 images, resulting in a 99% attack success rate.

**Fast Gradient Sign Method**

Definition: The FGSM [11] is an adversarial attack that creates perturbations to an input by taking a single step in the direction of the gradient of the loss function with respect to the input.

Parameters of the Attack:

- estimator – A trained classifier.
- norm – The norm of the adversarial perturbation. Possible values: "inf", np.inf, 1 or 2.
- eps – Attack step size (input variation).
- eps_step – Step size of input variation for minimal perturbation computation.
- targeted (bool) – Indicates whether the attack is targeted (True) or untargeted (False)
- num_random_init (int) – Number of random initialisations within the epsilon ball. For random_init=0 starting at the original input.
- batch_size (int) – Size of the batch on which adversarial samples are generated.
- minimal (bool) – Indicates if computing the minimal perturbation (True). If True, also define *eps_step* for the step size and eps for the maximum perturbation.

- summary_writer – Activate summary writer for TensorBoard. Default is *False* and deactivated summary writer. If *True* save runs/CURRENT_DATETIME_HOSTNAME in current directory. If of type *str* save in path. If of type *SummaryWriter* apply provided custom summary writer. Use hierarchical folder structure to compare between runs easily. e.g. pass in 'runs/exp1', 'runs/exp2', etc. for each new experiment to compare across them.

The following table outlines the default and modified parameters used in our ART experiments.

| Attack Parameter | Default Value | Manual Value |
| --- | --- | --- |
| Estimator | *CLASSIFIER_LOSS_GRADIENTS_TYPE* | TensorFlowV2Classifier |
| norm | Infinity | Infinity |
| eps | 0.3 | 0.03 |
| eps_step | 0.1 | 0.0005 |
| targeted | False | False |
| num_random_init | 0 | 0 |
| batch_size | 32 | 32 |
| minimal | False | False |
| summary_writer | False | False |

The effectiveness of these adjustments is demonstrated in the images below, showcasing the results of both the initial and modified attacks.

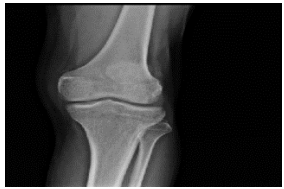

*Original Image*

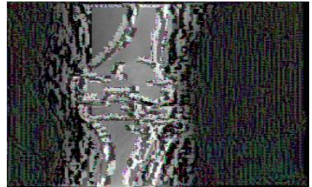

*Adversarial Image generated using FGSM attack by passing default parameters.*

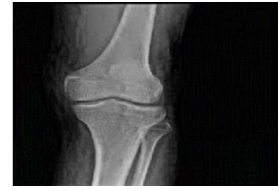

*Adversarial image generated using FGSM attack by passing altered parameters.*

**Attack Effectiveness**:

The FGSM attack was applied to 100 sample images. The above adversarial image on default parameters clearly depicts, there is so much noise in-comparison with adversarial image on altered parameters. The mean perturbation introduced was 0.017536785. Notably, this level of perturbation was sufficient to successfully generate adversarial samples for almost all 100 images, resulting in a 95% attack success rate.

## Simple Black-box Adversarial

Definition: A simple black-box adversarial attack [11] is a method of generating adversarial examples by querying a target model with various inputs to observe its predictions. The attacker uses these outputs to infer information about the model's decision boundaries and creates adversarial inputs that can mislead the model into making incorrect classifications.

Parameters of the Attack:

- classifier – A trained classifier predicting probabilities and not logits.
- attack (str) – attack type: pixel (px) or DCT (dct) attacks.
- max_iter (int) – The maximum number of iterations.

- epsilon (float) – Overshoot parameter.
- order (str) – order of pixel attacks: random or diagonal (diag).
- freq_dim (int) – dimensionality of 2D frequency space (DCT).
- stride (int) – stride for block order (DCT).
- targeted (bool) – perform targeted attack
- batch_size (int) – Batch size (but, batch process unavailable in this implementation).
- verbose (bool) – Show progress bars.

The following table outlines the default and modified parameters used in our ART experiments.

| Attack Parameter | Default Value | Manual Value |
| --- | --- | --- |
| Classifier | *CLASSIFIER_TYPE* | TensorFlowV2Classifier |
| attack | *dct* | *dct* |
| max_iter | 3000 | 3000 |
| epsilon | 0.1 | 0.125 |
| order | *random* | *random* |
| freq_dim | 4 | 4 |
| stride | 1 | 1 |
| targeted | False | False |
| batch_size | 1 | 1 |
| verbose | True | True |

The effectiveness of these adjustments is demonstrated in the images below, showcasing the results of both the initial and modified attacks.

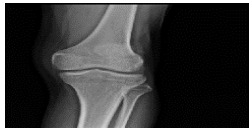

*Original Image*

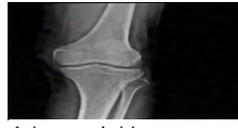

*Adversarial Image generated using SIMBA attack by passing default parameters.*

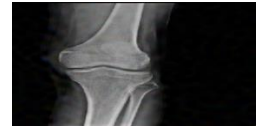

*Adversarial image generated using SIMBA attack by passing altered parameters.*

**Attack Effectiveness**:

The Simba attack was applied to 100 sample images. The above adversarial image on altered parameters has relatively more noise than comparison to adversarial image on default parameters. The mean perturbation introduced was 0.007455054. Notably, this level of perturbation was sufficient to successfully generate adversarial samples for almost all 100 images, resulting in a 47% attack success rate.

**Carlini-Wagner L2**

Definition: The Carlini-Wagner L2 [11] adversarial attack is an optimization-based method that aims to create adversarial examples by minimizing the L2 distance between the original input and the adversarial example while ensuring that the adversarial example is classified as a target class with high confidence.

Parameters of the Attack:

- classifier – A trained classifier.
- confidence (float) – Confidence of adversarial examples: a higher value produces examples that are farther away, from the original input, but classified with higher confidence as the target class.
- targeted (bool) – Should the attack target one specific class.

- learning_rate (float) – The initial learning rate for the attack algorithm. Smaller values produce better results but are slower to converge.
- binary_search_steps (int) – Number of times to adjust constant with binary search (positive value). If *binary_search_steps* is large, then the algorithm is not very sensitive to the value of *initial_const*. Note that the values gamma=0.999999 and c_upper=10e10 are hardcoded with the same values used by the authors of the method.
- max_iter (int) – The maximum number of iterations.
- initial_const (float) – The initial trade-off constant $c$ to use to tune the relative importance of distance and confidence. If *binary_search_steps* is large, the initial constant is not important, as discussed in Carlini and Wagner (2016).
- max_halving (int) – Maximum number of halving steps in the line search optimization.
- max_doubling (int) – Maximum number of doubling steps in the line search optimization.
- batch_size (int) – Size of the batch on which adversarial samples are generated.
- verbose (bool) – Show progress bars.

The following table outlines the default and modified parameters used in our ART experiments.

| Attack Parameter | Default Value | Manual Value |
|---|---|---|
| classifier | *CLASSIFIER_CLASS_LOSS_GRADIENTS_TYPE* | TensorFlowV2Classifier |
| confidence | 0.0 | 0.0 |
| targeted | False | False |
| learning_rate | 0.01 | 0.01 |
| binary_search_steps | 10 | 20 |
| max_iter | 10 | 10 |
| initial_const | 0.01 | 0.1 |
| max_halving | 5 | 5 |
| max_doubling | 5 | 5 |
| batch_size | 1 | 1 |
| verbose | True | True |

The effectiveness of these adjustments is demonstrated in the images below, showcasing the results of both the initial and modified attacks.

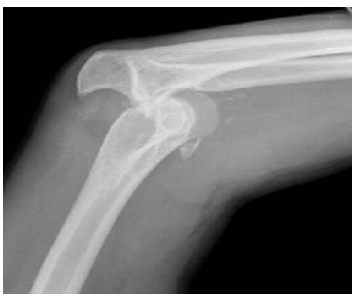

*Original Image*

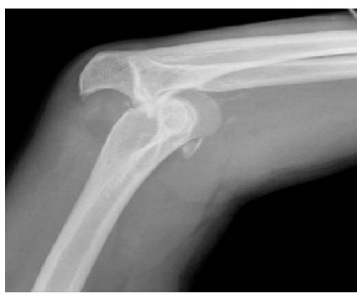

*Adversarial Image generated using Carlini-Wagner L2 attack by passing default parameters.*

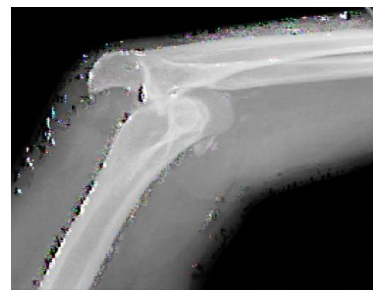

*Adversarial image generated using Carlini-Wagner L2 attack by passing altered parameters.*

**Attack Effectiveness:**

The Carlini-Wagner L2 attack was applied to 100 sample images. The above adversarial image on altered parameters has relatively more noise than comparison to adversarial image on default parameters. The mean perturbation introduced was 0.0040231724. Notably, this level of perturbation was sufficient to successfully generate adversarial samples for almost all 100 images, resulting in a 83% attack success rates.

**DeepFool**

Definition: The DeepFool [11] attack is an adversarial attack that approximates the decision boundaries of a classifier by treating them as linear hyperplanes in the vicinity of the input. It calculates the smallest perturbation required to change the classification of an input from its original class to the nearest decision boundary of a different class.

Parameters of the Attack:

- classifier – A trained classifier.
- max_iter (int) – The maximum number of iterations.
- epsilon (float) – Overshoot parameter.
- nb_grads (int) – The number of class gradients (top nb_grads w.r.t. prediction) to compute. This way only the most likely classes are considered, speeding up the computation.
- batch_size (int) – Batch size
- verbose (bool) – Show progress bars.

The following table outlines the default and modified parameters used in our ART experiments.

| Attack Parameter | Default Value | Manual Value |
| --- | --- | --- |
| Classifier | *CLASSIFIER_CLASS_LOSS_GRADIENTS_TYPE* | TensorFlowV2Classifier |
| max_iter | 100 | 500 |
| epsilon | *1e-06* | *1e-100* |
| nb_grads | 10 | 10 |
| batch_size | 1 | 1 |
| verbose | True | True |

The effectiveness of these adjustments is demonstrated in the images below, showcasing the results of both the initial and modified attacks.

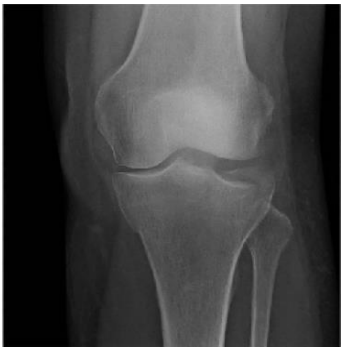

*Original Image*

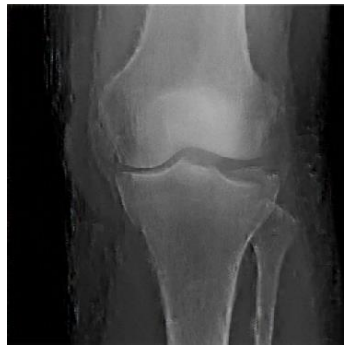

*Adversarial Image generated using DeepFool attack by passing default parameters.*

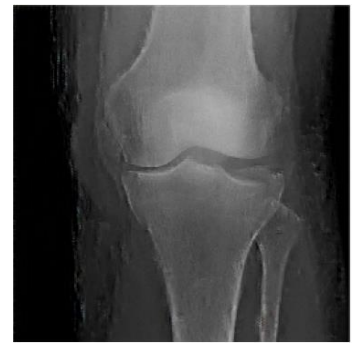

*Adversarial image generated using DeepFool attack by passing altered parameters.*

**Attack Effectiveness:**

The DeepFool attack was applied to 100 sample images. The above adversarial image on altered parameters has relatively less noise than adversarial image on default parameters. The mean perturbation introduced was 0.09733131. Notably, this level of perturbation was sufficient to successfully generate adversarial samples for almost all 100 images, resulting in a 99% attack success rate.

**Pixel**

Definition: A pixel [11] attack is an adversarial attack method that generates adversarial examples by altering pixel values in an input image. The goal is to create small, targeted changes that are often imperceptible to human observers while significantly affecting the model's predictions.

Parameters of the Attack:

- classifier – A trained classifier.
- th – threshold value of the Pixel/ Threshold attack. th=None indicates finding a minimum threshold.
- es (int) – Indicates whether the attack uses CMAES (0) or DE (1) as Evolutionary Strategy.
- max_iter (int) – Sets the Maximum iterations to run the Evolutionary Strategies for optimisation.
- targeted (bool) – Indicates whether the attack is targeted (True) or untargeted (False).
- verbose (bool) – Indicates whether to print verbose messages of ES used.

The following table outlines the default and modified parameters used in our ART experiments.

| Attack Parameter | Default Value | Manual Value |
|---|---|---|
| Classifier | *CLASSIFIER_NEURALNETWORK_TYPE* | TensorFlowV2Classifier |
| th | None | None |
| es | *1* | *1* |
| max_iter | 100 | 10 |
| targeted | False | False |
| verbose | False | False |

The effectiveness of these adjustments is demonstrated in the images below, showcasing the results of both the initial and modified attacks.

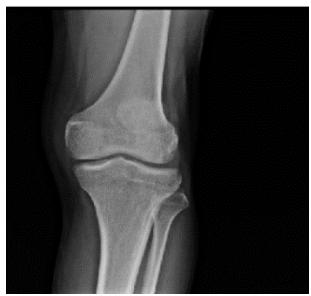

*Original Image*

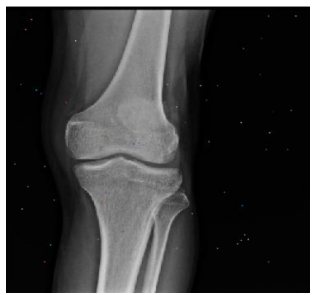

*Adversarial Image generated using pixel attack by passing default parameters.*

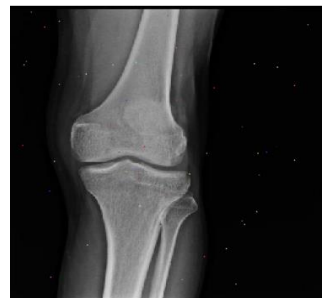

*Adversarial image generated using pixel attack by passing altered parameters.*

**Attack Effectiveness:**

The Pixel attack was applied to 100 sample images. The above adversarial image on altered parameters has almost same noise as adversarial image on default parameters. The mean perturbation introduced was 0.00018541123. Notably, this level of perturbation was sufficient to successfully generate adversarial samples for almost all 100 images, resulting in a 61% attack success rate.

**Evaluation of Adversarial Attack Efficacy for Bone Break Use case**

| Attack | Mean Perturbation |
|---|---|
| SIMBA | 0.007455054 |
| DeepFool | 0.09733131 |
| FGSM | 0.017536785 |
| Pixel | 0.00018541123 |
| Basic Iterative Method | 0.018666986 |
| Carlini-Wagner L2 Attack | 0.0040231724 |

To quantify the subtlety of adversarial attacks, we measure mean perturbation, which reflects the average magnitude of pixel modifications needed to induce misclassification.

While all attacks in this study employed subtle pixel-wise modifications, the Pixel Attack method stood out for its minimal level of perturbation. This characteristic makes Pixel Attack a particularly effective technique for generating adversarial examples that are difficult to detect.

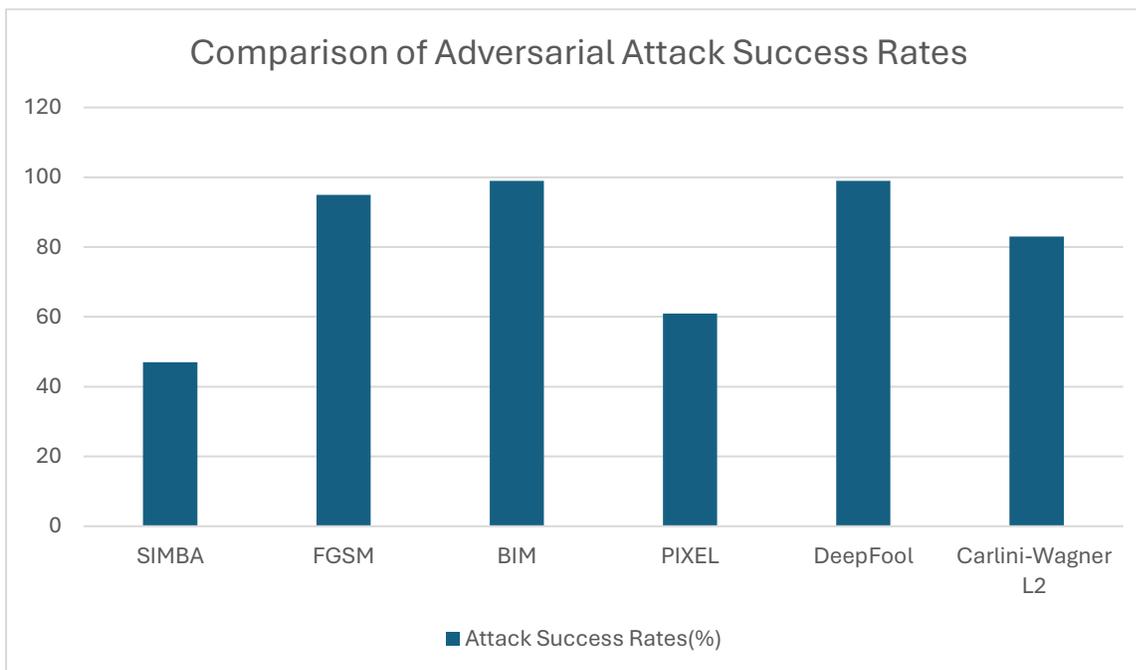

## Conclusion for attacks on Bone Break use case:

Our research underscores the critical security implications of adversarial attacks on CNN models. We have demonstrated the ease with which adversarial examples can be generated, potentially compromising the reliability of CNN-based systems in real-world applications. This highlights the urgent need for robust defenses to safeguard these systems from malicious exploitation.

Our findings underscore the critical vulnerability of AI models to adversarial attacks, where even subtle modifications can lead to significant performance degradation, especially in safety-critical applications. While existing defense techniques can offer some level of protection, our research highlights the urgent need for continuous innovation to stay ahead of evolving attack methods and ensure the robustness of AI systems.

## Defense Techniques and Evaluation

Following the application of adversarial attacks on images and the analysis of their detrimental effects, we now turn our attention to potential solutions for mitigating these threats. Various strategies exist to address adversarial attacks, including preprocessing inputs prior to model inference, training models with adversarial examples, and implementing adversarial input detection mechanisms. In this paper, we specifically focus on defense strategies that preprocess incoming inputs before they are fed into the model. This approach allows us to avoid the need for retraining the original model, thereby preserving its accuracy, while alternative defense mechanisms will be explored in future research endeavors by Responsible AI Office.

From our earlier analysis, we identified which adversarial attacks are particularly harmful and effective. To mitigate these threats, we implemented seven preprocessing defense techniques: Gaussian Blur, Gaussian Smoothing/Filtering, Feature Squeezing, Spatial Smoothing, Bilateral Filtering, Total Variation Denoising, and Median Filtering. We assessed the efficacy of each method against specific types of attacks, conducting experiments with adversarial samples generated by each attack technique and providing a detailed comparative analysis. Each defense strategy will be discussed in relation to its corresponding attack, evaluating their effectiveness based on the success rate of restoring original predictions. This success rate will be quantified as the percentage of adversarial images correctly classified after applying the defense strategy, relative to the total number of adversarial images tested. Each method is explored in detail in the following sections, highlighting their mechanisms and effectiveness against adversarial attacks.

## Evaluation Metrics

To evaluate the performance of the defense strategies, we employed the following metrics:

- **Success Rate:** The percentage of adversarial examples that were correctly classified by the model after applying the defense.
- **Perturbation Mean:** The average difference in pixel values between the original image and the adversarial image after applying the defense. A smaller value indicates a more effective defense.
- **Peak Signal-to-Noise Ratio (PSNR):** PSNR is a common metric used to measure the quality of reconstruction of a signal, especially in image and video processing. It's a ratio between the maximum possible power of a signal and the power of corrupting noise that affects the fidelity of its representation. A higher PSNR value generally indicates a better-quality reconstruction.

## Experimental Setup

To implement the defense mechanisms, we leveraged various Python libraries and tools. For median filtering and Gaussian smoothing, we utilized the SciPy library. OpenCV was employed for bilateral filtering and Gaussian blur. To incorporate more advanced techniques like spatial smoothing and feature squeezing, we utilized the Adversarial Robustness Toolbox (ART) framework. By effectively combining these tools, we were able to apply the defense mechanisms to the generated adversarial examples and assess their impact on model performance.

We will delve into defense strategies for both image classification and object detection models, discussing each use case separately.

## Defenses on Image Classification Use Case

### Median Filtering

Median filtering is a non-linear digital image processing technique used to reduce noise in images. It involves replacing each pixel in an image with the median value of its neighboring pixels. This operation effectively removes impulsive noise, such as salt-and-pepper noise, which introduces random white or black pixels into an image.

Here are the results obtained when applying Median filtering defense to images generated by various adversarial attacks.

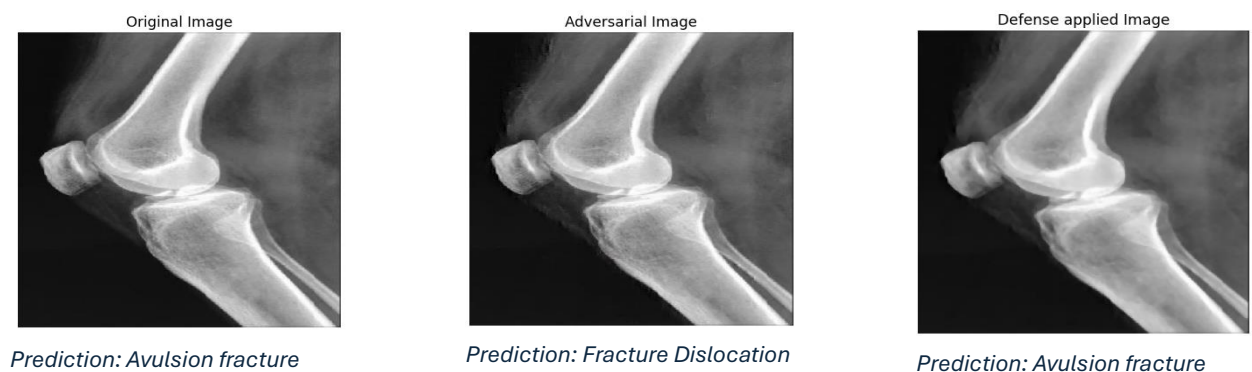

*Prediction: Avulsion fracture*

*Prediction: Fracture Dislocation*

*Prediction: Avulsion fracture*

Performance Metrics of Median filtering Defense Method Against Adversarial Attacks

| Attack | Defense Success Rate(%) on Adversarial Image. | Perturbation Mean between original image and adversarial image. | Perturbation mean between original and defense applied adversarial image. | PSNR between original and defense applied adversarial image. |
|---|---|---|---|---|
| Simba | 34 | 0.009046639 | 0.011598125 | 83.02885099 |
| BasicIterativeMethod | 1 | 0.01614459 | 0.014618156 | 80.1379395390389 |
| DeepFool Attack | 9 | 0.0017662274 | 0.0067326566 | 84.376338672373 |
| FastGradientMethod | 1 | 0.016796261 | 0.021183314 | 78.3307943274988 |
| Carlini-Wagner L2 | 71 | 0.0019001351 | 0.0062839715 | 85.7378362964391 |
| Pixel | 98 | 0.00030887345 | 0.0046982686 | 86.7732048770077 |

When median filtering was applied to adversarial images generated by all types of attacks considered in this research, it achieved a success rate of 35.7%. This means that 35.7% of the total adversarial images were able to retain their original predictions after the application of this defense, indicating its effectiveness in mitigating the noise and perturbations introduced by the adversarial attack.

### Total Variation Denoising

Total Variation (TV) denoising is a powerful image processing technique used to remove noise while preserving edges. It is based on the principle of minimizing the total variation of the image, which is a measure of the overall smoothness of the image. By minimizing the total variation, TV denoising effectively reduces noise while preserving sharp edges and fine details.

Here are the results obtained when applying Total Variation (TV) denoising defense to images generated by various adversarial attacks.

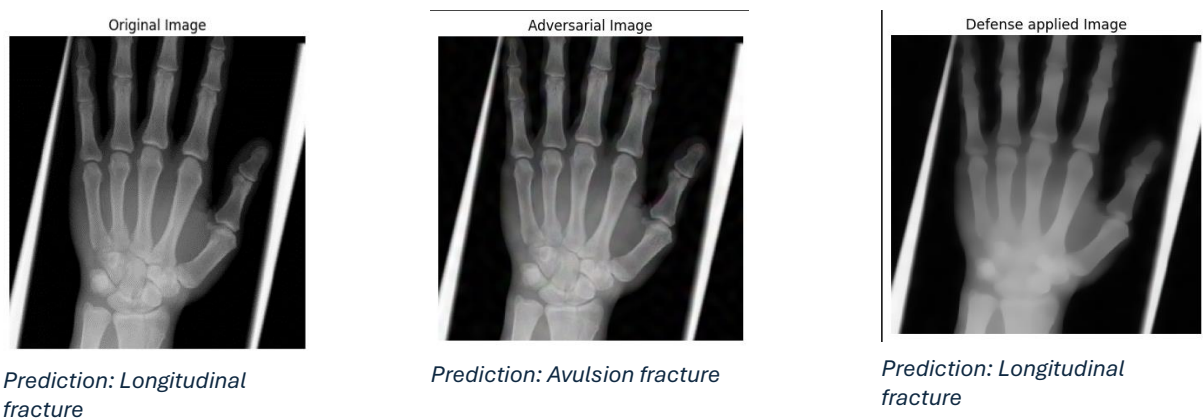

Prediction: Longitudinal fracture

Prediction: Avulsion fracture

Prediction: Longitudinal fracture

Performance Metrics of Total Variation (TV) denoising Defense Method Against Adversarial Attacks

| Attack | Defense Success Rate(%) on Adversarial Image. | Perturbation Mean between original image and adversarial image. | Perturbation mean between original and defense applied adversarial image. | PSNR between original and defense applied adversarial image. |
|---|---|---|---|---|
| Simba | 36 | 0.008892781 | 0.014371217 | 81.7723382801442 |
| BasicIterativeMethod | 17 | 0.019168012 | 0.02683921 | 75.2808045515885 |
| DeepFool Attack | 8 | 0.0013686273 | 0.013817731 | 81.3575201222088 |
| FastGradientMethod | 22 | 0.016428167 | 0.015569179 | 81.0696471920862 |
| Carlini-Wagner L2 | 83 | 0.0019599 | 0.012184532 | 82.2894378415734 |
| Pixel | 90 | 0.00032761367 | 0.010996273 | 82.5023235198159 |

When Total Variation (TV) denoising was applied to adversarial images generated by all types of attacks considered in this research, it achieved a success rate of 42.7%. This means that 42.7% of the total adversarial images were able to retain their original predictions after the application of this defense, indicating its effectiveness in mitigating the noise and perturbations introduced by the adversarial attack.

**Bilateral Filtering**

Bilateral filtering is a non-linear image processing technique used to remove noise while preserving edges. It combines spatial filtering, which considers the proximity of pixels, with range filtering, which considers the similarity of pixel values. This approach allows bilateral filtering to effectively remove noise while preserving important image features such as edges and textures.

Here are the results obtained when applying Bilateral filtering defense to images generated by various adversarial attacks.

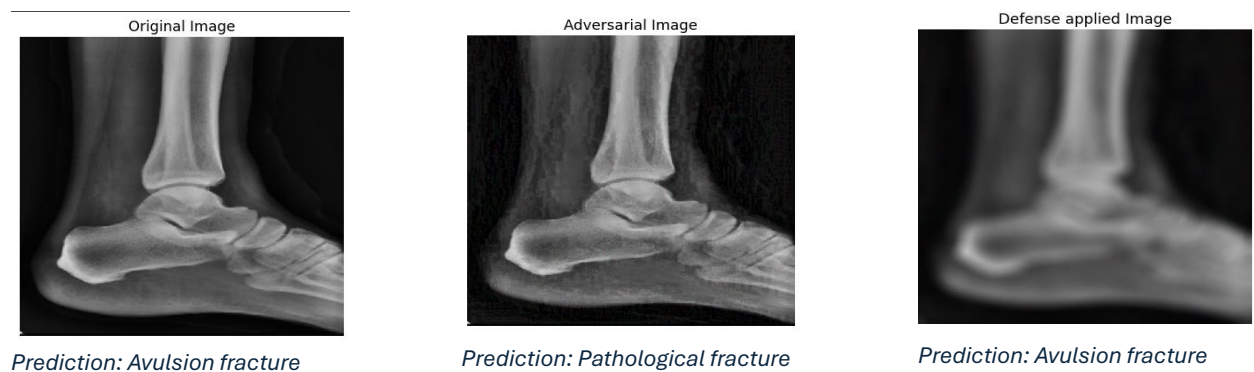

*Prediction: Avulsion fracture*          *Prediction: Pathological fracture*          *Prediction: Avulsion fracture*

Performance Metrics of Bilateral filtering Defense Method Against Adversarial Attacks

| Attack | Defense Success Rate(%) on Adversarial Image. | Perturbation Mean between original image and adversarial image. | Perturbation mean between original and defense applied adversarial image. | PSNR between original and defense applied adversarial image. |
|---|---|---|---|---|
| **Simba** | 6 | 0.006272632 | 0.026387341 | 75.35094168 |
| **BasicIterativeMethod** | 19 | 0.018896846 | 0.026538573 | 75.1499460438771 |
| **DeepFool Attack** | 8 | 0.0045292173 | 0.022948822 | 75.5186890097438 |
| **FastGradientMethod** | 34 | 0.017707312 | 0.02540432 | 75.532807055411 |
| **Carlini-Wagner L2** | 49 | 0.0024668432 | 0.022328721 | 75.7180191302525 |
| **Pixel** | 52 | 0.00032078972 | 0.02074727 | 76.1654834620397 |

When Bilateral filtering was applied to adversarial images generated by all types of attacks considered in this research, it achieved a success rate of 28%. This means that 28% of the total adversarial images were able to retain their original predictions after the application of this defense, indicating its effectiveness in mitigating the noise and perturbations introduced by the adversarial attack.

**Spatial Smoothing**

Spatial smoothing is a general term that encompasses various techniques used to reduce noise and artifacts in images by applying filtering operations to the spatial dimensions of the image. These techniques typically involve averaging or combining the values of neighboring pixels to create a smoother image.

Here are the results obtained when applying Spatial smoothing defense to images generated by various adversarial attacks.

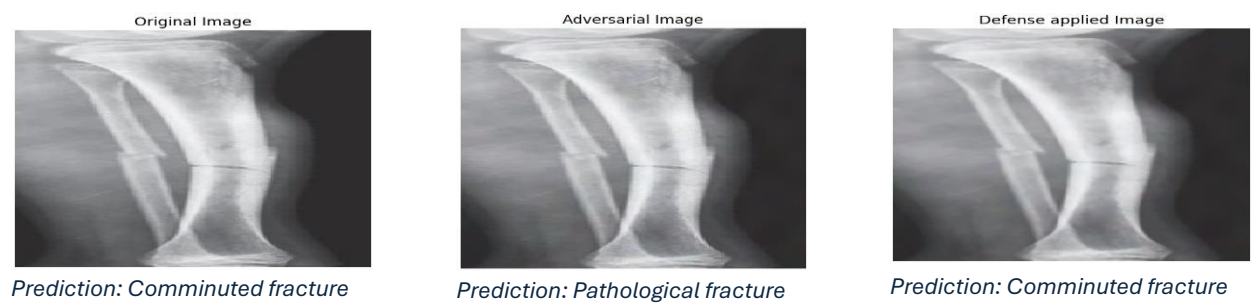

*Prediction: Comminuted fracture*          *Prediction: Pathological fracture*          *Prediction: Comminuted fracture*

Performance Metrics of Spatial smoothing Defense Method Against Adversarial Attacks.

| Attack | Defense Success Rate(%) on Adversarial Image. | Perturbation Mean between original image and adversarial image. | Perturbation mean between original and defense applied adversarial image. | PSNR between original and defense applied adversarial image. |
|---|---|---|---|---|
| **Simba** | 25 | 0.00892589 | 0.012473643 | 82.3856378172374 |
| **BasicIterativeMethod** | 1 | 0.01614459 | 0.015143435 | 80.0187471007733 |
| **DeepFool Attack** | 8 | 0.0015133036 | 0.006719814 | 84.0766205389275 |
| **FastGradientMethod** | 1 | 0.016796261 | 0.021443868 | 78.2209928593678 |
| **Carlini-Wagner L2** | 70 | 0.001907365 | 0.0062567433 | 85.6729755139432 |
| **Pixel** | 98 | 0.00030887345 | 0.0046500843 | 86.782548284985 |

When Spatial smoothing was applied to adversarial images generated by all types of attacks considered in this research, it achieved a success rate of 33.8%. This means that 33.8% of the total adversarial images were able to retain their original predictions after the application of this defense, indicating its effectiveness in mitigating the noise and perturbations introduced by the adversarial attack.

**Feature Squeezing**

Feature squeezing is a technique used to reduce the dimensionality of input data while preserving important information. It involves applying transformations that compress the data into a lower-dimensional space, making it easier to process and analyze.

Here are the results obtained when applying Feature squeezing defense to images generated by various adversarial attacks, as evaluated using our evaluation matrix.

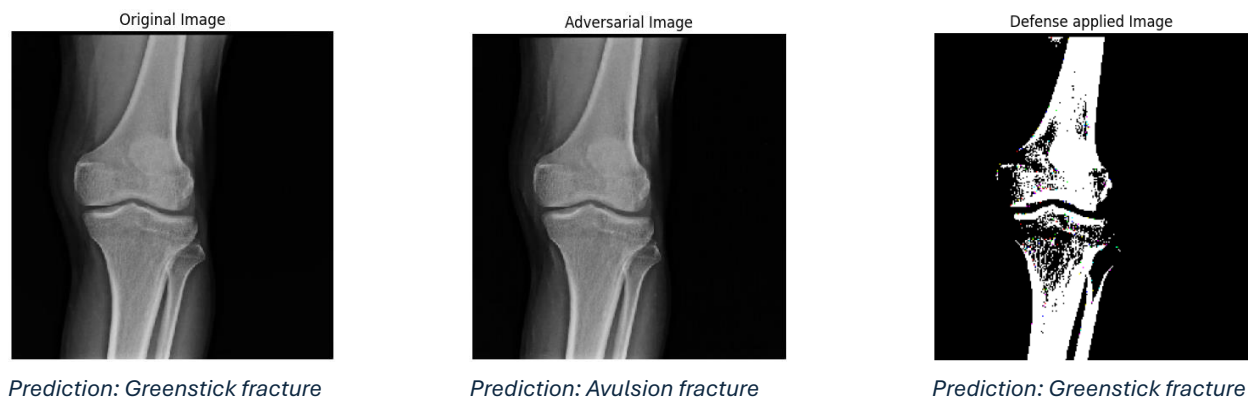

Original Image — *Prediction: Greenstick fracture*

Adversarial Image — *Prediction: Avulsion fracture*

Defense applied Image — *Prediction: Greenstick fracture*

Performance Metrics of Feature squeezing Defense Method Against Adversarial Attacks.

| Attack | Defense Success Rate(%) on Adversarial Image. | Perturbation Mean between original image and adversarial image. | Perturbation mean between original and defense applied adversarial image. | PSNR between original and defense applied adversarial image. |
|---|---|---|---|---|
| **Simba** | 17 | 0.009974818 | 0.23227806 | 58.9696767153886 |
| **BasicIterativeMethod** | 8 | 0.017508954 | 0.23155646 | 58.9226153283354 |
| **DeepFool Attack** | 4 | 0.018207898 | 0.24723476 | 58.66415238 |
| **FastGradientMethod** | 13 | 0.017757215 | 0.24396318 | 58.6423494855692 |
| **Carlini-Wagner L2** | 24 | 0.0038535867 | 0.24811158 | 58.5967925943198 |
| **Pixel** | 18 | 0.00035948362 | 0.21468309 | 59.2111894606979 |

When Feature squeezing was applied to adversarial images generated by all types of attacks considered in this research, it achieved a success rate of 14%. This means that 14% of the total adversarial images were able to retain their original predictions after the application of this defense, indicating its effectiveness in mitigating the noise and perturbations introduced by the adversarial attack.

## Gaussian smoothing/filtering

Gaussian smoothing is a convolution operation that applies a Gaussian kernel to an image. The Gaussian kernel is a two-dimensional function that has a bell-shaped curve. The width of the Gaussian kernel determines the degree of smoothing applied to the image.

Here are the results obtained when applying Gaussian smoothing defense to images generated by various adversarial attacks, as evaluated using our evaluation matrix.

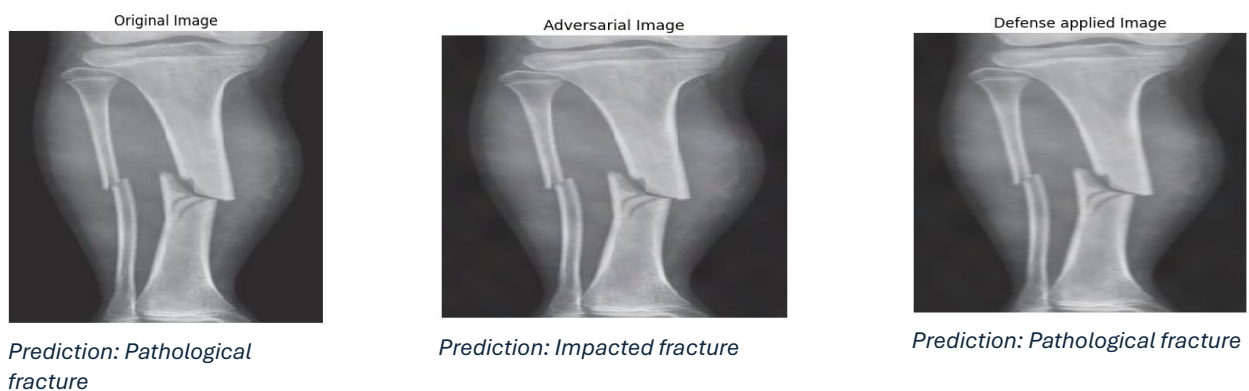

Original Image

*Prediction: Pathological fracture*

Adversarial Image

*Prediction: Impacted fracture*

Defense applied Image

*Prediction: Pathological fracture*

Performance Metrics of Gaussian smoothing Defense Method Against Adversarial Attacks.

| Attack | Defense Success Rate(%) on Adversarial Image. | Perturbation Mean between original image and adversarial image. | Perturbation mean between original and defense applied adversarial image. | PSNR between original and defense applied adversarial image. |
|---|---|---|---|---|
| **Simba** | 17 | 0.008819914 | 0.010282724 | 85.0026978918619 |
| **BasicIterativeMethod** | 1 | 0.01614459 | 0.016832082 | 80.0416627633768 |
| **DeepFool Attack** | 5 | 0.0013560762 | 0.006691057 | 86.4723275873822 |
| **FastGradientMethod** | 7 | 0.018670322 | 0.01720408 | 79.7212334996212 |
| **Carlini-Wagner L2** | 57 | 0.0015902037 | 0.0058634565 | 87.0215396056212 |
| **Pixel** | 92 | 0.00032917812 | 0.0053591346 | 86.008803327084 |

When Gaussian smoothing was applied to adversarial images generated by all types of attacks considered in this research, it achieved a success rate of 29.8%. This means that 29.8% of the total adversarial images were able to retain their original predictions after the application of this defense, indicating its effectiveness in mitigating the noise and perturbations introduced by the adversarial attack.

**Gaussian blur**

Gaussian blur is a type of image processing technique that applies a Gaussian kernel to an image, resulting in a blurred or softened version of the original. This process is often used to reduce noise, smooth edges, and create a more visually appealing image.

Here are the results obtained when applying Gaussian blur denoising defense to images generated by various adversarial attacks.

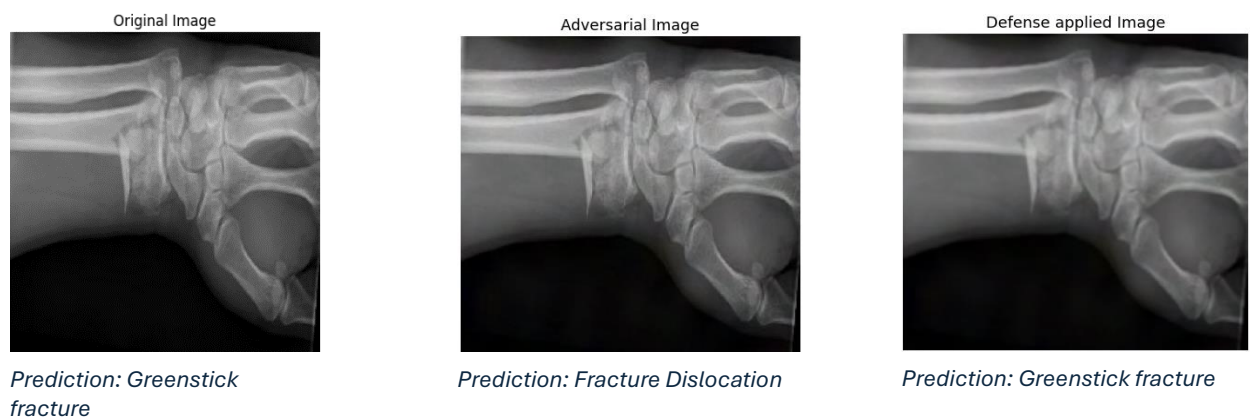

Original Image

Adversarial Image

Defense applied Image

*Prediction: Greenstick fracture*

*Prediction: Fracture Dislocation*

*Prediction: Greenstick fracture*

Performance Metrics of Gaussian smoothing Defense Method Against Adversarial Attacks.

| Attack | Defense Success Rate(%) on Adversarial Image. | Perturbation Mean between original image and adversarial image. | Perturbation mean between original and defense applied adversarial image. | PSNR between original and defense applied adversarial image. |
|---|---|---|---|---|
| **Simba** | 17 | 0.0074760155 | 0.012132031 | 82.6067884352161 |
| **BasicIterativeMethod** | 1 | 0.01614459 | 0.015107942 | 81.7745467362285 |
| **DeepFool Attack** | 7 | 0.0010344564 | 0.0074425843 | 84.6917277521117 |
| **FastGradientMethod** | 2 | 0.017807506 | 0.014924283 | 81.855474540852 |
| **Carlini-Wagner L2** | 58 | 0.0015440531 | 0.0066927685 | 85.210579612189 |
| **Pixel** | 97 | 0.00031387398 | 0.0062242867 | 84.710688874761 |

When Gaussian blur was applied to adversarial images generated by all types of attacks considered in this research, it achieved a success rate of 30.3%. This means that 30.3% of the total adversarial images were able to retain their original predictions after the application of this defense, indicating its effectiveness in mitigating the noise and perturbations introduced by the adversarial attack.

While traditional defenses often proved insufficient against the evolving threat of adversarial attacks, we explored defensive distillation.

**Defensive Distillation**

Defensive distillation is a technique that enhances the robustness of neural networks against adversarial attacks. By training a smaller, simpler network (the "student") to mimic the output probabilities of a larger, more complex network (the "teacher"), the student becomes less susceptible to adversarial perturbations. This process involves training the student on soft targets, which are probability distributions over all classes generated by the teacher. Soft targets lead to smoother gradients during training, reducing the impact of small input changes on the network's output. This, in turn, makes it harder for attackers to craft effective adversarial examples that can mislead the network.

To explore the effectiveness of defensive distillation, we trained a model using the ART library on a limited dataset of 100 original samples. This experimental setup allowed us to assess the potential benefits of the technique in a controlled research environment.

Here are the results obtained when applying Defensive distillation defense to images generated by various adversarial attacks.

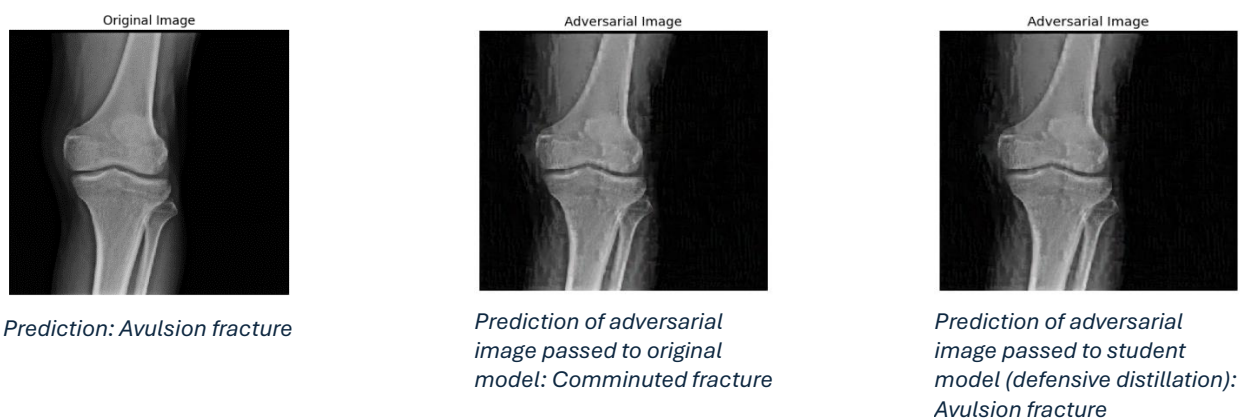

*Prediction: Avulsion fracture*

*Prediction of adversarial image passed to original model: Comminuted fracture*

*Prediction of adversarial image passed to student model (defensive distillation): Avulsion fracture*

Performance Metrics of Defensive distillation Defense Method Against Adversarial Attacks.

| Attack | Defense Success Rate(%) on Adversarial Image. |
|---|---|
| **Simba** | 94 |
| **BasicIterativeMethod** | 79 |
| **DeepFool Attack** | 46 |
| **FastGradientMethod** | 57 |
| **Carlini-Wagner L2** | 90 |
| **Pixel** | 62 |

The implementation of defensive distillation yielded promising results in mitigating adversarial attacks. The model demonstrated a 71% success rate against a diverse range of adversarial attacks considered in this research, highlighting its effectiveness in enhancing the robustness of the neural network.

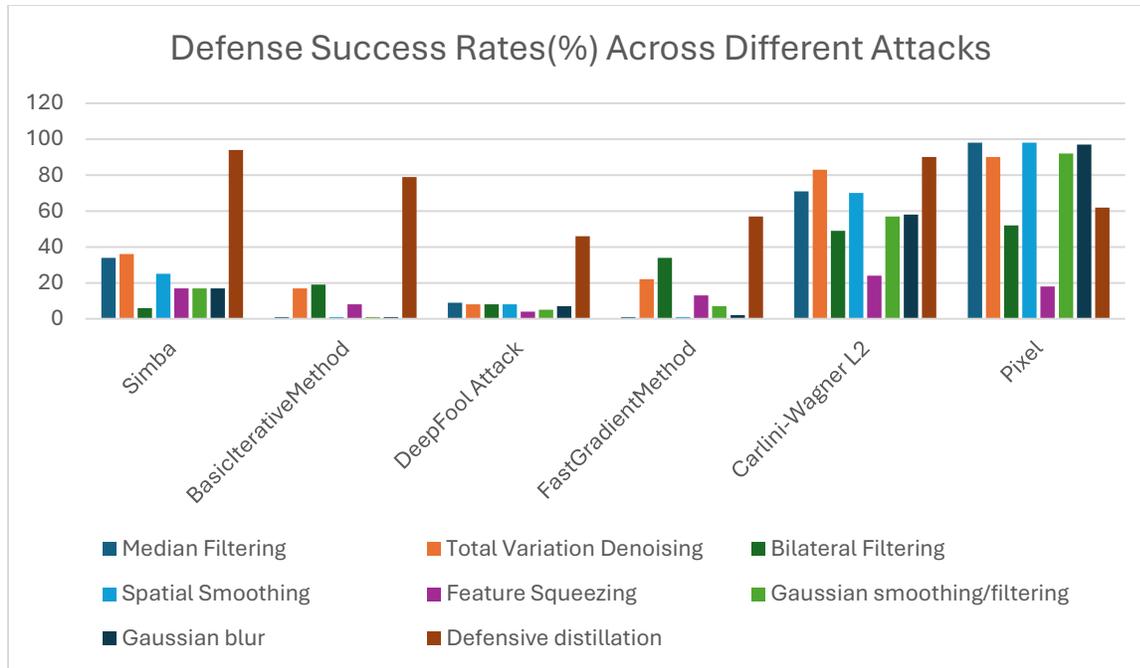

**Conclusion of Defenses on Image Classification Use Case:**

Our experimental evaluation demonstrated the efficacy of defensive distillation in mitigating adversarial attacks against CNN models. While preprocessing techniques proved to be less effective, defensive distillation exhibited promising results, achieving a success rate of approximately 71%. This highlights the potential of defensive distillation as a robust defense mechanism against adversarial attacks, which can easily deceive models with minimal perturbations.

### Defenses on Object Detection Use Case

While we successfully applied various defense mechanisms to mitigate adversarial attacks on image classification models, we did not attempt defensive distillation due to its high computational cost associated with training a student model. This is an avenue we may explore in future research at our Responsible AI Office.

In our experiments with object detection model, we found that preprocessing techniques were ineffective in restoring original predictions after adversarial attacks, yielding success rates below 1%. This highlights the necessity for developing novel defense mechanisms specifically tailored to object detection, which we aim to pursue in our upcoming research efforts.

**Conclusion**

This research investigated the vulnerability of AI models, specifically CNNs and object detection models, to adversarial attacks. By leveraging the ART library, we successfully generated adversarial examples that could easily deceive these models, highlighting the critical need for robust defense mechanisms. While preprocessing techniques proved ineffective in mitigating these attacks, defensive distillation demonstrated promising results in enhancing the robustness of image classification models. However, we did not explore defensive distillation for object detection models due to its high computational cost. These findings underscore the urgency of developing novel and effective defense strategies to safeguard AI systems from adversarial threats. Future research will focus on exploring advanced defense techniques and addressing the unique challenges posed by adversarial attacks on object detection models.

## Limitations

While our research provides valuable insights into the vulnerability of AI models to adversarial attacks and the potential of defense mechanisms, several limitations need to be acknowledged:

1. **Limited Dataset Size**: Our experiments were conducted on a relatively small dataset of 100 samples. A larger and more diverse dataset would provide a more comprehensive evaluation of the effectiveness of defense techniques.
2. **Potential Impact on Model Performance**: Defensive distillation, while effective in mitigating adversarial attacks, may compromise the original model's performance. Further research is needed to optimize the training process and minimize the performance trade-off.
3. **Evolving Attack Techniques**: The landscape of adversarial attacks is constantly evolving. More sophisticated attacks can be crafted by exploiting specific vulnerabilities in model architectures. Therefore, it is crucial to continuously develop and refine defense strategies to stay ahead of these threats.
4. **Limitations of Preprocessing Defenses**: Preprocessing techniques, such as those employed in our research, can be effective in mitigating certain types of attacks. However, they may not be sufficient against more advanced attacks that target the decision boundaries of the model.
5. **Computational Cost of Defense Mechanisms**: Some defense techniques, such as defensive distillation, can be computationally expensive, especially for large models. Exploring efficient implementations and hardware acceleration techniques is essential for practical deployment.

## Learnings and Future Work

Our research provided valuable insights into the vulnerability of AI models to adversarial attacks and the limitations of existing defense mechanisms. Key learnings include:

- **Ease of Adversarial Attack Creation:** We demonstrated the ease with which adversarial examples can be generated using black-box attack techniques, even without detailed knowledge of the model architecture.
- **Inefficacy of Preprocessing Defenses:** Preprocessing techniques, while simple to implement, proved to be insufficient in mitigating the impact of adversarial attacks.
- **Potential of Defensive Distillation:** Defensive distillation emerged as a promising defense strategy, particularly for image classification models. However, its effectiveness for object detection models is yet to be tested.

Future research directions should focus on:

- **Advanced Defense Techniques:** Exploring more sophisticated defense mechanisms, such as adversarial training, certified robustness, and input transformation techniques.
- **Domain-Specific Defenses:** Developing defense strategies tailored to specific domains like autonomous vehicles, healthcare, and finance, where the consequences of adversarial attacks can be severe.
- **Adversarial Training with Diverse Attacks:** Incorporating a diverse set of adversarial attacks into the training process to improve model robustness.
- **Real-world Deployment Considerations:** Addressing the challenges of deploying defense mechanisms in real-world scenarios, including computational cost, latency, and compatibility with different hardware platforms.

By addressing these research directions, we can strive to develop more robust and resilient AI systems that are capable of withstanding adversarial attacks.